\documentclass{aa}
\usepackage{amsmath}
\usepackage{txfonts}
\usepackage{graphicx}
\usepackage{natbib}
\usepackage{amsfonts}
\usepackage{amssymb}
\usepackage{hyperref}
\usepackage{longtable}
\usepackage{color}

\input cyracc.def
\font\tencyr=wncyr10
\def\cyr{\tencyr\cyracc}

\bibpunct{(}{)}{;}{a}{}{,} 

\renewcommand{\hat}[1]{\oldhat{\mathbf{#1}}}

\begin{document}
\title{Impact on asteroseismic analyses of regular gaps in \emph{Kepler} data}  
\author{R.~A.~Garc\'\i a\inst{1} \and 
S. Mathur\inst{2, 1}\and
S. Pires\inst{1}\and
C. R\'egulo\inst{3,4}\and
 B. Bellamy\inst{5}\and
 P.L. Pall\'e\inst{3,4}\and
 J.~Ballot \inst{6,7} \and
 S. Barcel\'o Forteza\inst{3,4}\and
 P.G. Beck\inst{8}\and
T.R. Bedding\inst{5}\and
T. Ceillier\inst{1}\and
T. Roca Cort\'es \inst{3,4}\and
D. Salabert\inst{1}\and
D.~Stello \inst{5}
}
 \institute{Laboratoire AIM, CEA/DSM -- CNRS - Univ. Paris Diderot -- IRFU/SAp, Centre de Saclay, 91191 Gif-sur-Yvette Cedex, France
\and Space Science Institute, 4750 Walnut Street, Suite 205, Boulder, Colorado 80301 USA
\and Instituto de Astrof\'isica de Canarias, E-38200 La Laguna, Tenerife, Spain
\and Departamento de Astrof\'isica,  Universidad de La Laguna, E-38206 La Laguna, Tenerife, Spain
\and Sydney Institute for Astronomy, School of Physics, University of Sydney, NSW 2006, Australia
\and Universit\'e de Toulouse, UPS-OMP, IRAP, 31400 Toulouse, France
\and CNRS, Institut de Recherche en Astrophysique et Plan\'etologie, 14 avenue Edouard Belin, 31400 Toulouse, France
\and Instituut voor Sterrenkunde, Katholieke Universiteit Leuven, Celestijnenlaan 200D, B-3001 Leuven, Belgium
}

\date{Received 01 November 2013/ Accepted }
\abstract{The NASA \emph{Kepler} mission has observed more than 190,000 stars in the constellations of Cygnus and Lyra. Around 4 years of almost continuous ultra high-precision photometry have been obtained reaching a duty cycle higher than 90\% for many of these stars. However,  almost regular gaps due to nominal operations are present in the light curves at different time scales.}
{In this paper we want to highlight the impact of those regular gaps in asteroseismic analyses and we try to find a method that minimizes their effect in the frequency domain.}
{To do so, we isolate the two main time scales of quasi regular gaps in the data. We then interpolate the gaps and we compare the power density spectra of four different stars: two red giants at different stages of their evolution, a young F-type star, and a classical pulsator in the instability strip. }
{The spectra obtained after filling the gaps in the selected solar-like stars show a net reduction in the overall background level, as well as a change in the background parameters. The inferred convective properties could change as much as  $\sim$200$\%$ in the selected example, introducing a bias in the p-mode frequency of maximum power. When global asteroseismic scaling relations are used, this bias can lead up to a variation in the surface gravity of 0.05 dex. Finally, the oscillation spectrum in the classical pulsator is cleaner compared to the original one.}
{}
\authorrunning{R.A. Garc\'\i a et al.}
\titlerunning{Impact of {\emph{Kepler}} nominal regular gaps in asteroseismic analyses}
\maketitle

\section{Introduction}

Since its launch until the end of the nominal mission (2009-2013) \emph{Kepler}  \citep{2010Sci...327..977B} has collected light curves for almost 200,000 stars with varied observation lengths ranging from one month to almost four years \citep[e.g.][]{2014ApJS..211....2H}. The telescope pointed towards the Cygnus \& Lyra constellations with a 115 deg$^2$ field of view. The mission has already accomplished many breakthroughs on extrasolar planet research \citep[e.g.][]{2010Sci...330...51H,2011ApJ...729...27B,2012ApJ...745..120B,2012Sci...337..556C,2013Sci...340..587B,2013Sci...342..331H}. In addition, it is contributing greatly to  stellar physics through asteroseismology. The long and quasi-continuous light curves of several years have made possible to probe global properties of thousands of solar-like stars and red giants \citep[][]{2011Sci...332..213C,2014ApJS..210....1C,2011ApJ...729L..10B,2012A&A...537A..30M,2013ApJ...765L..41S}, as well as their internal structure \citep[e.g.][]{2011Natur.471..608B,2012ApJ...749..152M,2013MNRAS.435..242G}, and their internal rotation \citep{2012Natur.481...55B,2012A&A...548A..10M,2012ApJ...756...19D}.


The Earth-trailing orbit of the mission helped to maximize the observing time while optimizing the duty cycle (in many cases above 90\%).  During its nominal mission, \emph{Kepler} underwent  several operations affecting the scientific data acquisition and producing almost regular gaps in the light curves. Although these gaps have a small impact on the exoplanet search in the time domain, they can have significant impact on seismic studies based on the frequency domain. 

\begin{figure}[htbp]
\begin{center}
\includegraphics[width=9.0cm]{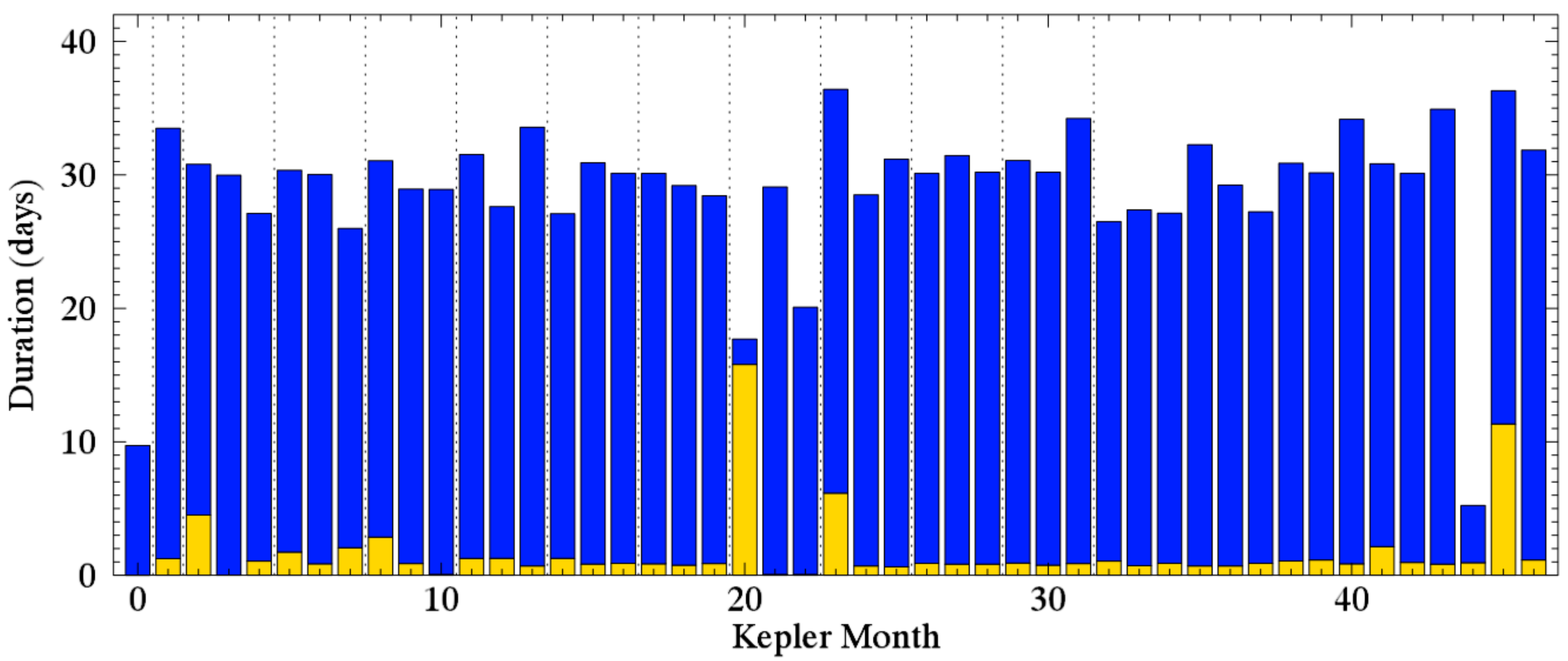}
\caption{Duration of observations (Blue bars) and data gaps (yellow bars) during the  47 ``\emph{Kepler} months'', from Q0 to Q16. The \emph{Kepler} scientific observations started on May 2, 2009 at 00:55 UT time.}
\label{Fig_bar}
\end{center}
\end{figure}

There were three main operations causing quasi-regular gaps in the \emph{Kepler} scientific data acquisition. The first was the Angular Momentum Dump (AMD), in which the reaction wheels --used for attitude control of the satellite-- were desaturated. In general, they occurred every $\sim$3 days producing typical gaps of one long-cadence (29.42 minutes) or several short-cadence (58.85 s) measurements \citep[for more details see the \emph{Kepler} Data Characteristics Handbook][hereafter KDCH]{Christiansen2013_data_handbook}. The second cause was the Downlink Earth Pointing (DEP). At the end of every ``\emph{Kepler} month'' (30.12 days, on average), the spacecraft stopped observing for $\sim$0.9 days to send the data stored in the satellite to ground (see Fig.~\ref{Fig_bar}).  Finally, a third regular operation was performed four times per year --once every Quarter (Q)--, when the satellite was rolled by $90^\circ$ about its axis to keep the solar panels illuminated and the radiators away from the Sun. This operation generally took place during the standard DEP and no additional time was required. However, three gaps were longer than the median value of the data interruptions (see Fig.~\ref{Fig_bar}). They were the gaps between Q1 and Q2, Q7 and Q8, and between the first and second month of Q16.  

{In addition to these regular events, there were other irregular instrumental problems producing small interruptions in the scientific operations, including the reaction wheels crossing zero angular momentum, losses of spacecraft fine pointing, and sudden pixel sensitivity drop-outs. These caused occasional outliers visible in the \emph{Kepler} Simple Aperture Photometry light curves (usually called the raw data series).} Due to the random nature of these events, the impact on the frequency spectrum was less noticeable than the aforementioned semi-regular events.

The main consequences of the regular gaps are to modify the background power in the Fourier domain, and to introduce sequences of harmonics at high frequency. It is therefore important to properly address this potential problem because it can bias  the estimations of the parameters of the convective background, and the value of $\nu_{\rm{max}}$, i.e., the frequency of the maximum power of the acoustic modes \citep[e.g.][]{2009CoAst.160...74H,2010A&A...511A..46M}. This latter parameter is used to infer global properties of stars such as their masses and radii \citep[e.g.][]{2008ApJ...674L..53S,2011Sci...332..213C,2011ApJ...741..119M,2011A&A...530A.142B,2014ApJS..210....1C}.

In this paper we start in section 2 by studying  the impact of the quasi-regular gaps in the \emph{Kepler} spectral window.  In section 3, we present a method to minimize the effects of such regular gaps and we show some results concerning the analysis of the convective background of a young F-type stars, the study of two evolved red giants at different stages of their evolution, and a typical classical pulsator in the instability strip. 

\section{Impact of the \emph{Kepler} window function}
\label{Sect2}

\begin{figure}[!htb]
\begin{center}
\includegraphics[width=9cm]{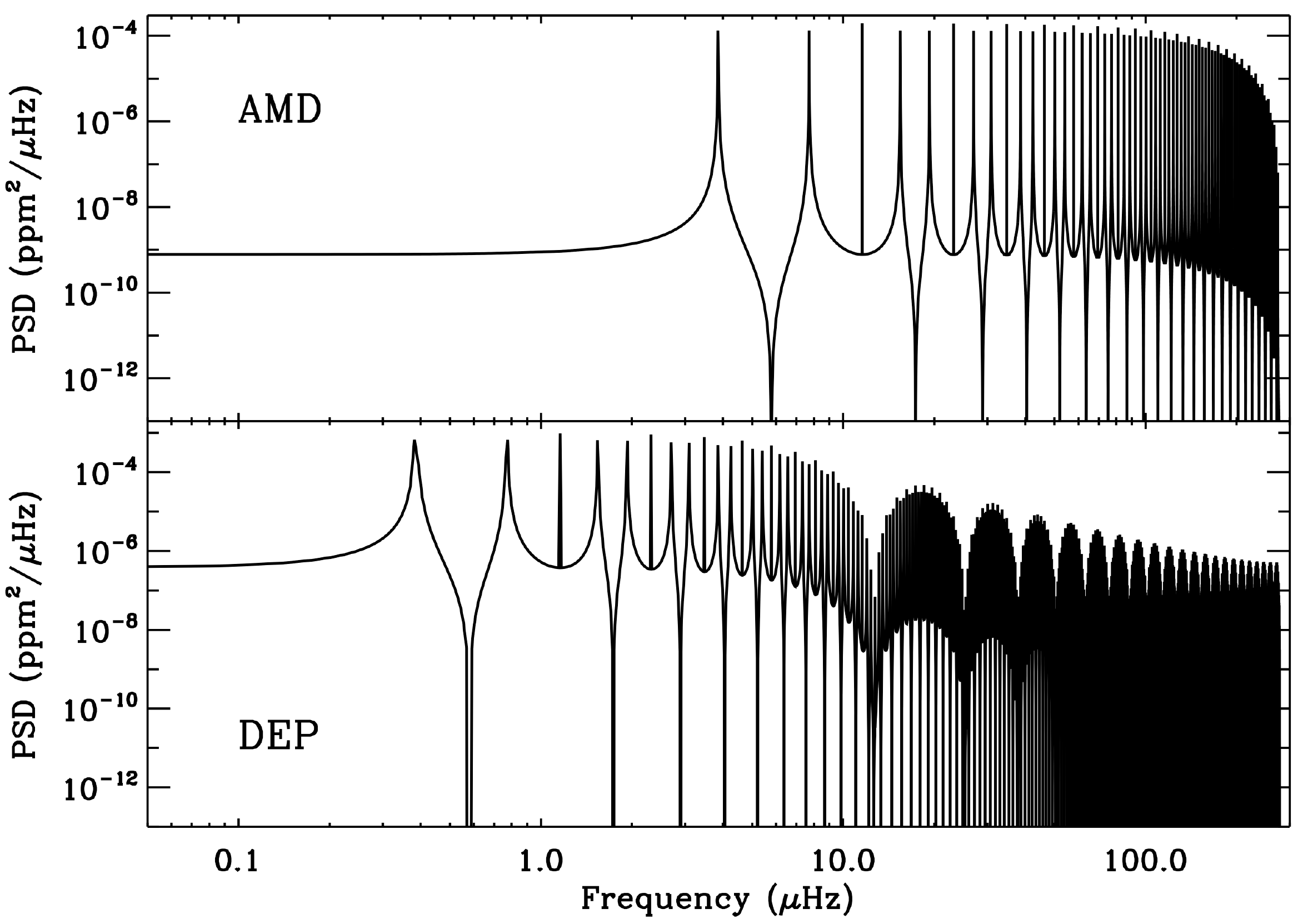}
\caption{Spectral windows of the gaps induced by the Angular Momentum Dump (AMD, top) and the Downlink Earth Pointing (DEP, bottom) respectively. }
\label{Fig_PSF}
\end{center}
\end{figure}

To characterize the impact of the interruption in the data acquisition due to the AMD and the DEP, we decompose the observed signal, $Y(t)$ as the ideal continuous signal, $X(t)$, multiplied by the window function, $W(t)$:
\begin{equation}
Y(t) = X(t)  \cdot W(t) \,,
\end{equation}
where the window function, $W(t)$, includes the two main \emph{Kepler} interruptions. In the case of long-cadence observations:
\begin{enumerate}
\item AMD gaps can be modeled by a Dirac Comb function:
\begin{equation}
W_1(t) =  {\mbox{\cyr Sh}}_{T_1}(t) \,,
\end{equation}
where $T_1$ is 3 days (see the spectral response in the top panel of Fig.~\ref{Fig_PSF}). 

\item The gaps due to the DEP can be modeled by a rectangular window convolved with a Dirac comb function:
\begin{equation}
W_2(t) =  \Pi_{T'_2}(t) \ast  {\mbox{\cyr Sh}}_{T_2}(t) \,,
\end{equation}
where $T'_2$ is 0.9 days and $T_2$ is 1 \emph{Kepler} month (see the spectral response in the bottom panel of Fig.~\ref{Fig_PSF}).

\end{enumerate}

On the one hand, the spectral window of the AMD gaps is dominated by a sequence of harmonics with the fundamental at a period of 3 days (3.86 $\mu$Hz). On the other hand, for the DEP gaps, the spectral response is more complex with a sequence of harmonics of 1 month period (0.39 $\mu$Hz) with smaller amplitudes towards high frequency. This variation of the amplitude could be important as it can modify the slope of the convective background in solar-like stars.



\begin{figure}[!htb]
\begin{center}
\includegraphics[width=9.0cm]{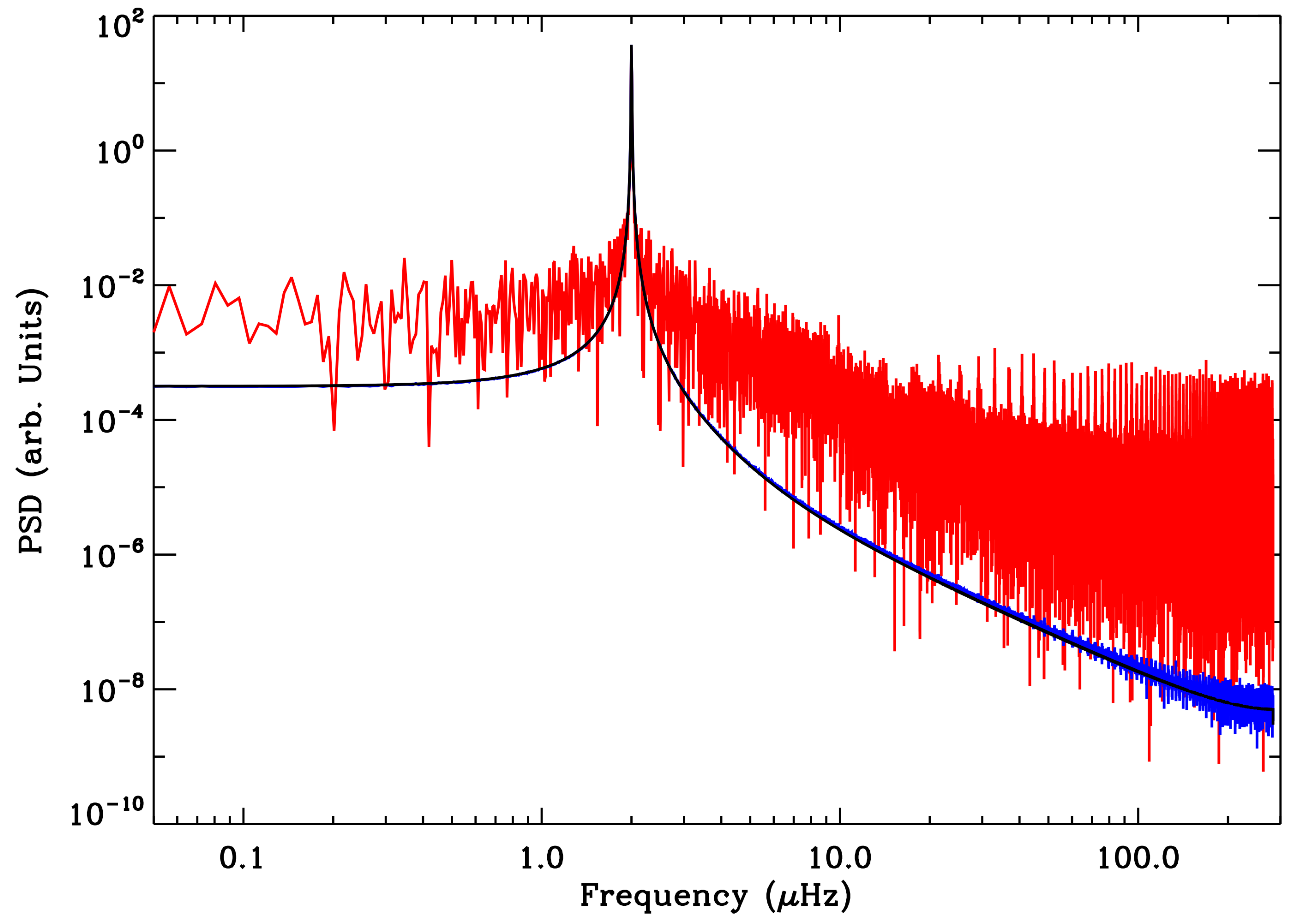}
\caption{Power Spectral Density (PSD) in logarithmic scale of a 2 $\mu$Hz ideal sine wave (black) following the Barycentric \emph{Kepler} Julian Date (BKJD, see Appendix~\ref{Ap_timing} for further details). The red curve corresponds to the PSD of the same sine wave multiplied by the \emph{Kepler} window function. Both PSDs were computed using a Lomb-Scargle algorithm. The blue line is the PSD resultant of converting the timing BKJD of the ideal sine wave (without gaps) into a regular grid and computed using a FFT.}
\label{Fig_SineWave}
\end{center}
\end{figure}

An example of the Power Spectral Density (PSD) of a sine wave of 2 $\mu$Hz frequency, computed using a Lomb-Scargle algorithm \citep{1982ApJ...263..835S}, is shown in Fig.~\ref{Fig_SineWave} (black curve). This sine wave has been simulated using the Barycentric \emph{Kepler} Julian Date (BKJD) corresponding to the long-cadence measurements of KIC~3733735 in the ideal case (without gaps). A detailed discussion about the \emph{Kepler} timing can be found in Appendix~\ref{Ap_timing}. 
The red curve corresponds to the PSD of the same simulation after we have multiplied the ideal observations by the \emph{Kepler} window function (i.e. adding the observational gaps). As expected, at high-frequency (above $\sim$ 10 $\mu$Hz) the spectrum is dominated by the 3-day harmonics due to the AMD gaps. The increase in the level of power is about 5 orders of magnitude close to the Nyquist frequency. At lower frequencies down to the central frequency of the sine wave, the spectrum shows an increase in the power compared to the ideal case without gaps, following the trends induced by the DEP gaps.

\section{Interpolating the gaps in  \emph{Kepler} observations}


In order to minimize the impact of the regular gaps in the \emph{Kepler} light curves, we propose to properly interpolate the time series. 

\begin{figure}[!htb]
\begin{center}
\includegraphics[width=9.0cm]{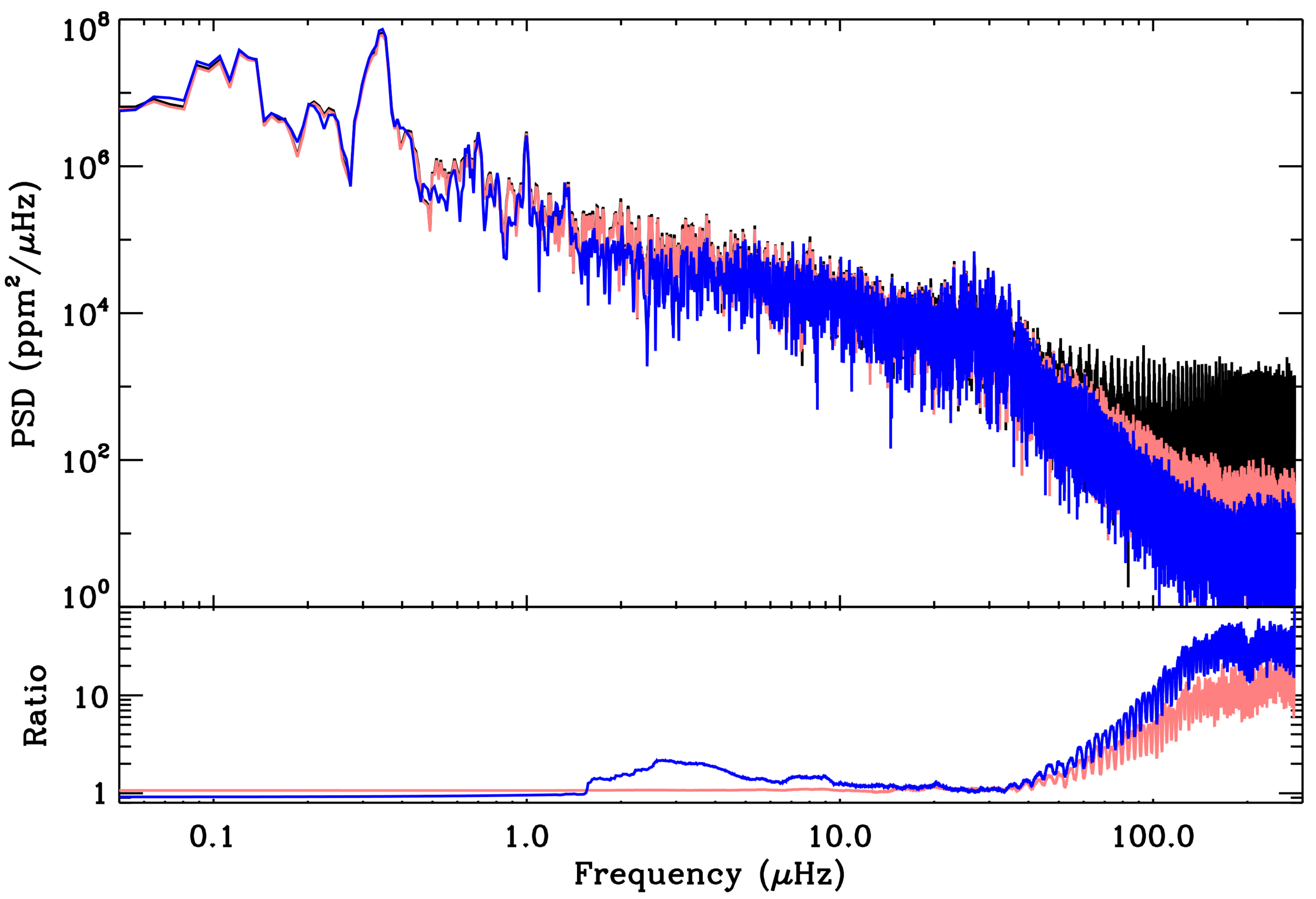}
\caption{Top: Power Spectral Density (PSD) in logarithmic scale of KIC~2305930 smoothed using a 3 point boxcar function. The black curve corresponds to the original light curve. The pink curve is the PSD after correcting all the gaps with sizes smaller than1 hour (two long-cadence points), while the blue curve is the PSD after inpainting all the gaps smaller than 20 days. Bottom: Ratio between the interpolated PSDs and the one with all the gaps. Both curves have been smoothed using a 300-point boxcar function.
}
\label{Fig_INP23}
\end{center}
\end{figure}

In the case of the AMD gaps --and because they are small (typically 1 long-cadence point every three days)-- the interpolation can be done with any  simple interpolation algorithm. To illustrate that, we compare the  PSD of a typical red giant observed in long cadence by \emph{Kepler} before and after interpolating the data. The Q0-Q16 { SAP} \citep[{ Simple Aperture Photometry,}][]{Christiansen2013_data_handbook} light curve of this star, KIC~2305930, has been corrected for  instrumental perturbations and the quarters have been stitched together following the procedures described by \citet{2011MNRAS.414L...6G}. The final light curve has been high-pass filtered using a triangular smooth function of a 100-day width. The PSD is dominated at high frequency by the harmonics of the AMD gaps (see black curve in Fig.\ref{Fig_INP23}).

{ It is important to note that the original NASA's  Pre-search Data Conditioning pipeline (PDC) --utilized to correct  the \emph{Kepler} light curves \citep{2010ApJ...713L.120J}-- high-pass filters the data such that periods larger than 3 days are slightly effected, while periods longer than 20 days are almost entirely removed
\citep[see Fig. 1 in][]{ThompsonRel21}. An improved version of this pipeline called PDC-MAP or PDC-msMAP --where msMAP stands for multi scale Maximum A Posteriori-- used a Bayesian approach to keep part of the low frequency signal. However, in some quarters the algorithm cannot distinguish between instrumental noise and stellar variability and the signal is completely removed for periods above $\sim$21 days \citep[for a comparison of the different corrections methods see][]{2013ASPC..479..129G}.  }

We have then interpolated all the gaps in the light curve of KIC~2305930 with lengths equal or smaller than one hour (hence only AMD gaps) using a $3^{\rm{rd}}$ order polynomial fit around the gaps \citep{2011A&A...530A..97B}. The resultant PSD is shown in Fig.~\ref{Fig_INP23} (pink curve). As expected, only the high-frequency harmonics of the AMD gaps are removed, the rest of the spectrum stays the same (see bottom panel of Fig.~\ref{Fig_INP23}). It is important to notice that the harmonics of the AMD gaps are only visible in the PSD if the star has periodic signals (e.g. surface rotation signatures or low-frequency pulsations) with periods longer than 3 days.

To go further, we need to interpolate the $\sim$ 0.9-day gaps between the ``\emph{Kepler} months'' due to the DEP --as well as the three longer ones-- in order to remove the spurious trends introduced by the missing data. For this purpose, it is necessary to use an interpolation algorithm that is able to deal with a large range of gap sizes while preserving the main oscillatory signal and the background as much as possible.

\begin{figure}[!htbp]
\begin{center}
\includegraphics[width=9.0cm]{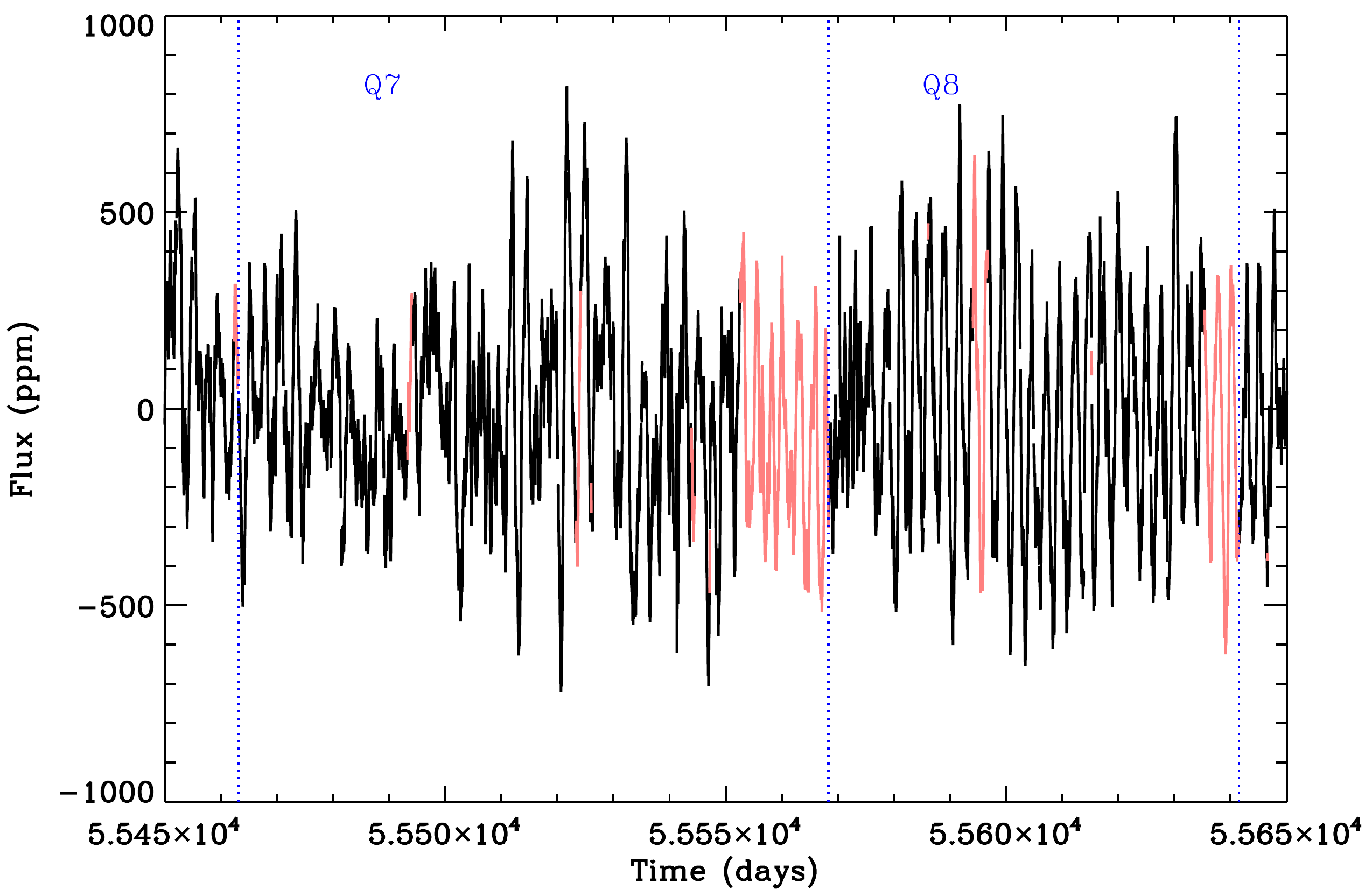}
\caption{Long cadence light curve of KIC~3733735 (black) for quarters Q7 and Q8. The limit of the quarters are indicated by vertical blue dotted lines. The red line corresponds to the inpainted segments.}
\label{Fig_inpaint}
\end{center}
\end{figure}

Several methods have been proposed to interpolate seismic data \citep[e.g.][]{1982MNRAS.199...53F,1990ApJ...349..667B}. Although they work quite well, they require an a-priori knowledge of the signal to be treated. Therefore, these techniques are not suited to treat thousands of unknown asteroseismic targets. We propose to use an inpainting technique \citep{Elad05} based on a prior of sparsity instead. This method assumes that there is a representation of the signal in which most of the coefficients are close to zero. For example, if the signal was a single sine wave, the sparsest representation would be the Fourier transform because most of the Fourier coefficients would be zero except one (hence sparse), which is sufficient to represent the sine wave in the frequency space. Therefore in asteroseismology, and to deal with the large variation of gap sizes (from 1 short-cadence data point to $\sim$16 days), the best representation is a Multi-scale Discrete Cosine Transform  \citep[MDCT,][]{2006aida.book.....S}. Inpainting techniques have already been used several times in astrophysics \citep[e.g.][]{2008StMet...5..289A,2009MNRAS.395.1265P}, as well as in asteroseismology to correct a few solar-like stars \citep[e.g.][]{2010A&A...518A..53M,2013A&A...549A..12M,2010arXiv1003.5178S,2011A&A...530A..97B} observed by the CoRoT satellite \citep{2006cosp...36.3749B}.
 { An example of the inpainted light curve of KIC~3733735 can be seen in Fig.~\ref{Fig_inpaint} for quarters Q7 and Q8.}

This inpainting algorithm has the additional advantage of being very fast compared to other techniques such as the CLEAN periodogram \citep{1987AJ.....93..968R,1995AJ....109.1889F}. It requires less than a minute for a Q0-Q16 long cadence light curve (see Pires et al. 2014 { for a more complete description of the algorithm and its application to asteroseismic data\footnote{The \emph{Kepler}-inpainting software can be found in \url{http://irfu.cea.fr/Sap/en/Phocea/Vie_des_labos/Ast/ast_visu.php?id_ast=3346}}}) when it is applied to regularly sampled data. In the case of \emph{Kepler}, we need to convert the BKJD irregular time series into a regular grid of points. To do so we use the nearest neighbor resampling algorithm \citep{TIM.2008.2009201}. Hence, we built a new time series with a sampling rate equal to the median of the original. Each point of the new series is built from the closest observation (irregularly sampled) if it is within half of the new sampling distance. The regular grid point is left empty (zero) if there is no original observation falling within the new sampling. An example of this methodology is shown in the blue curve of Fig.~\ref{Fig_SineWave}. The increase of noise is only visible above $\sim$~50~$\mu$Hz and it is negligible compared to the noise introduced by the window function (see Appendix~\ref{Ap_interpol} for further details).

Therefore, we propose to inpaint all the gaps in the \emph{Kepler} time series --with sizes smaller than 20 days-- into a regular grid of points with a sampling rate equal to the median sampling rate of each star. By doing so, the Nyquist frequency is fixed to the value determined by the regular sampling rate and we lost the possibility of exploring frequency regimes above the formal Nyquist frequency as recently done by \citet{2013MNRAS.430.2986M} and \citet{2013arXiv1312.4500B}. Finally, for gaps longer than 20 days --including missing quarters-- are too long to be treated in this way. { The inpainting algorithm needs to be tested further in these conditions which is out of the scope of this paper. }


The resultant power spectrum of the inpainted time series for KIC~2305930 is shown in Fig.~\ref{Fig_INP23} (hearafter all the PSDs plotted in blue are computed using inpainted time series). In this case, we see not only an improvement at high frequency, but also at intermediate frequencies, between 1.5 and 10 $\mu$Hz. Indeed the background level is reduced by a factor of $\sim$~2.5 in this region while the ratio was flat when only AMD gaps were interpolated (see bottom panel of Fig.~\ref{Fig_INP23}). In the region dominated by the p modes (10-40 $\mu$Hz), the PSD is the same (ratio of 1) and there is no significant improvement when we  interpolate the DEP gaps.

Because the effect of the \emph{Kepler} window function depends on the characteristics of the stellar signal, we also show the result of inpainting all the gaps in three other stars: KIC~8905990, an evolved M giant \citep[e.g.][]{2013MNRAS.436.1576B,2013A&A...559A.137M} with $\nu_{\rm{max}}\sim0.4$ $\mu$Hz observed in long cadence for the full mission; KIC~3733735, a young F-type star ($\nu_{\rm{max}}\sim 2130$ $\mu$Hz) continuously observed in short cadence since Q5 \citep[see e.g.][]{2012A&A...543A..54A,2014A&A...562A.124M}; and KIC~5892969, a classical pulsator in the instability strip.

\begin{figure}[!htbp]
\begin{center}
\includegraphics[width=9.0cm]{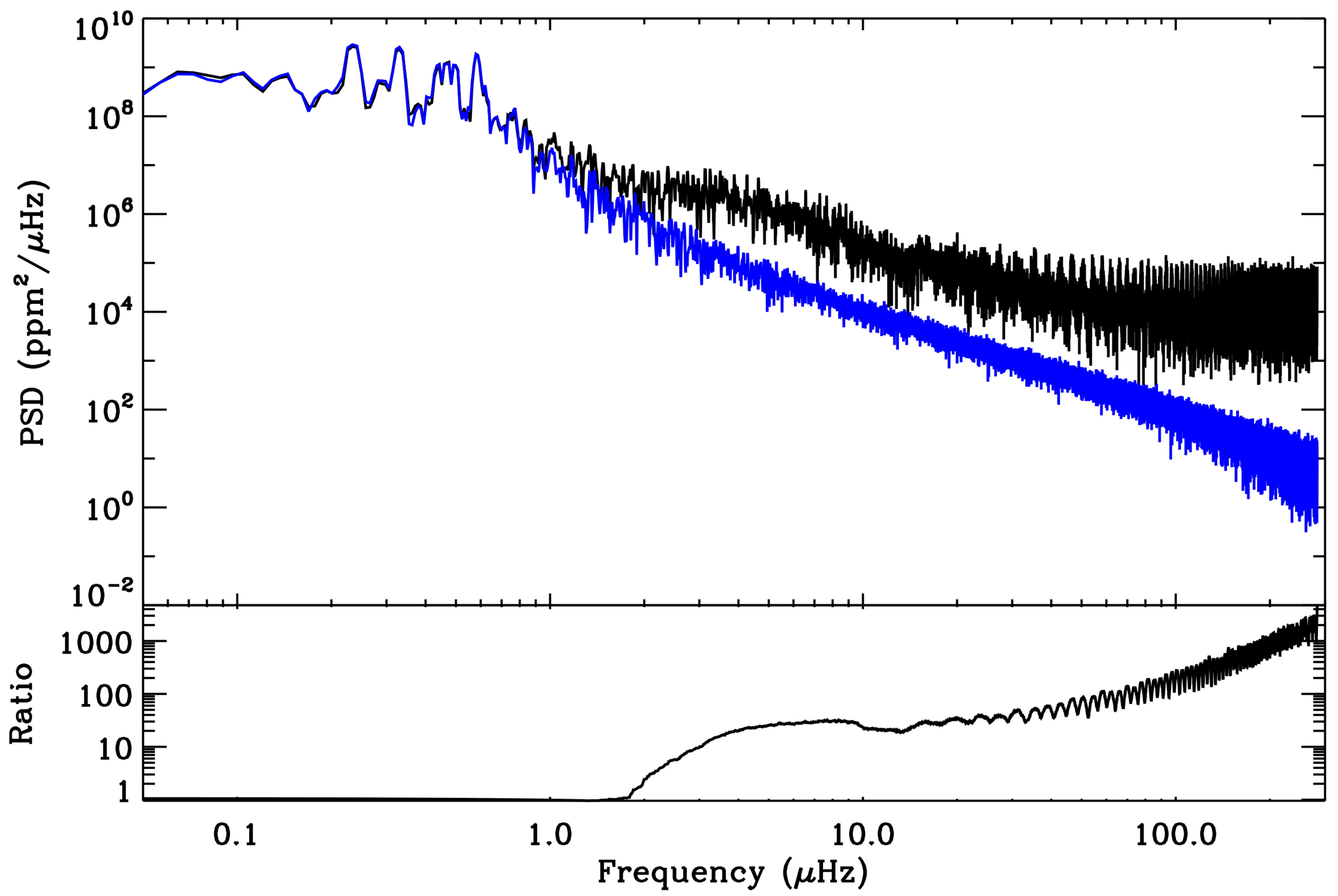}
\caption{As in Fig.~\ref{Fig_INP23}, but for the evolved red giant  KIC~8905990. In black the PSD for the original \emph{Kepler} time series, while in blue is the inpainted one. Bottom panel: ratio between both PSDs.}
\label{Fig_Result1}
\end{center}
\end{figure}

In the case of the evolved red giant KIC~8905990, the PSD of the long-cadence Q0-Q16 light curve corrected following \citet{2011MNRAS.414L...6G} is shown before and after inpainting the gaps in Fig.~\ref{Fig_Result1}. The improvement in the background  at frequencies above 2 $\mu$Hz is larger than in the previous case. The shape of the background was completely dominated by the spectral window (black curve) and after interpolating the gaps (blue curve), we reach a reduction in the noise level of 3 orders of magnitude at high frequency (see bottom panel of Fig.~\ref{Fig_Result1}). 

\begin{figure}[!htbp]
\begin{center}
\includegraphics[width=9.0cm]{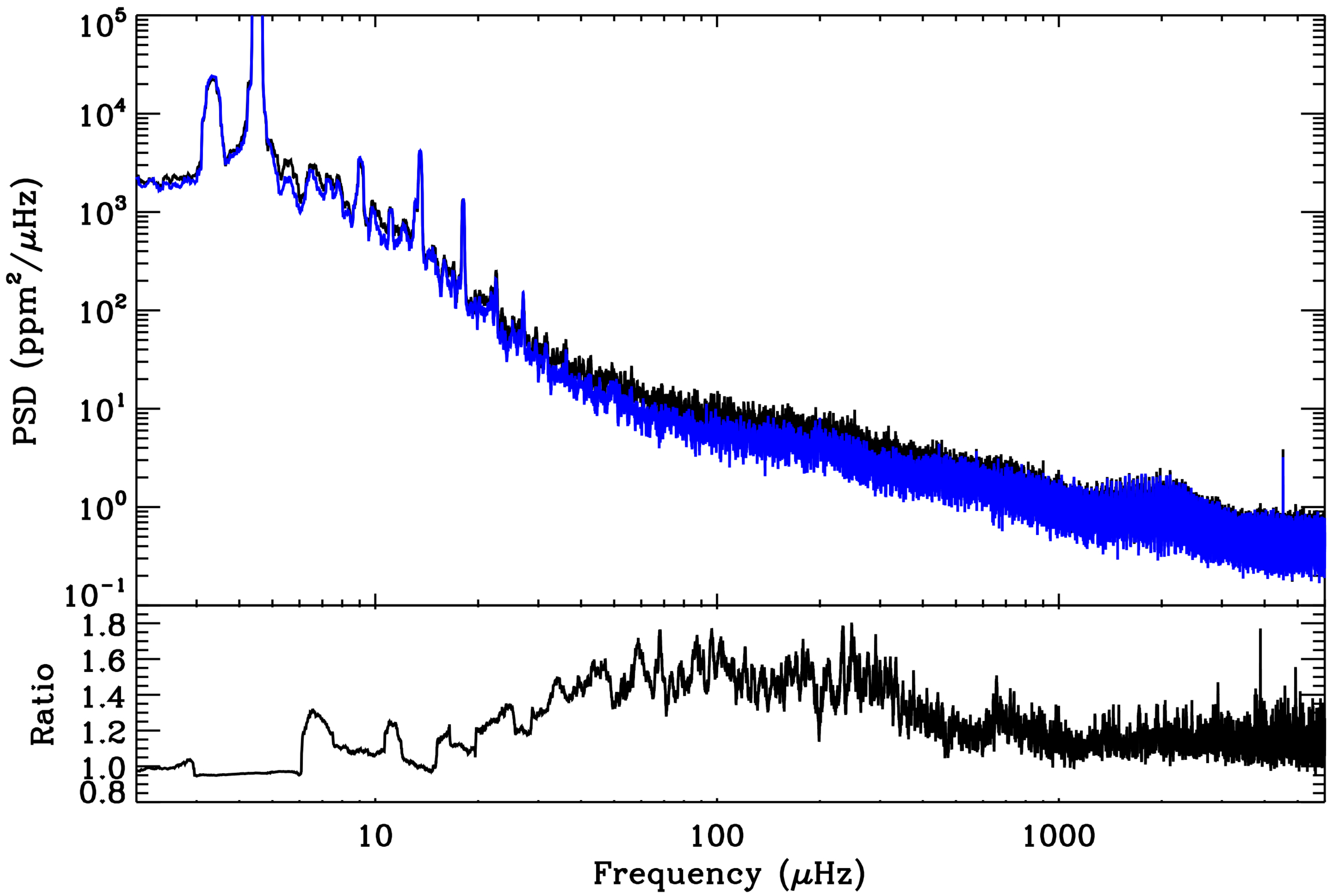}
\caption{Same than Fig.~\ref{Fig_Result1} but for the young F star KIC~3733735.}
\label{Fig_Result2}
\end{center}
\end{figure}

For the young F star KIC~3733735 observed in short cadence, we have used the Q5-Q16 PDC-msMAP \citep[Pre-search Data Conditioning-multi scale Maximum A Posteriori,][]{Christiansen2013_data_handbook} corrected data to show that the interpolation works well with any type of corrected light curves. The improvement in the inpainted PSD is located at high frequencies compared to the previous stars (see Fig.~\ref{Fig_Result2}). We achieve a reduction in the background of a factor of 1.6  between $\sim$50 and 400 $\mu$Hz. This region is dominated by the convective background and the reduction in the noise has a non negligible impact on the convection properties we can infer. 

\begin{figure*}[!htbp]
\begin{center}
\includegraphics[width=6cm, angle=90]{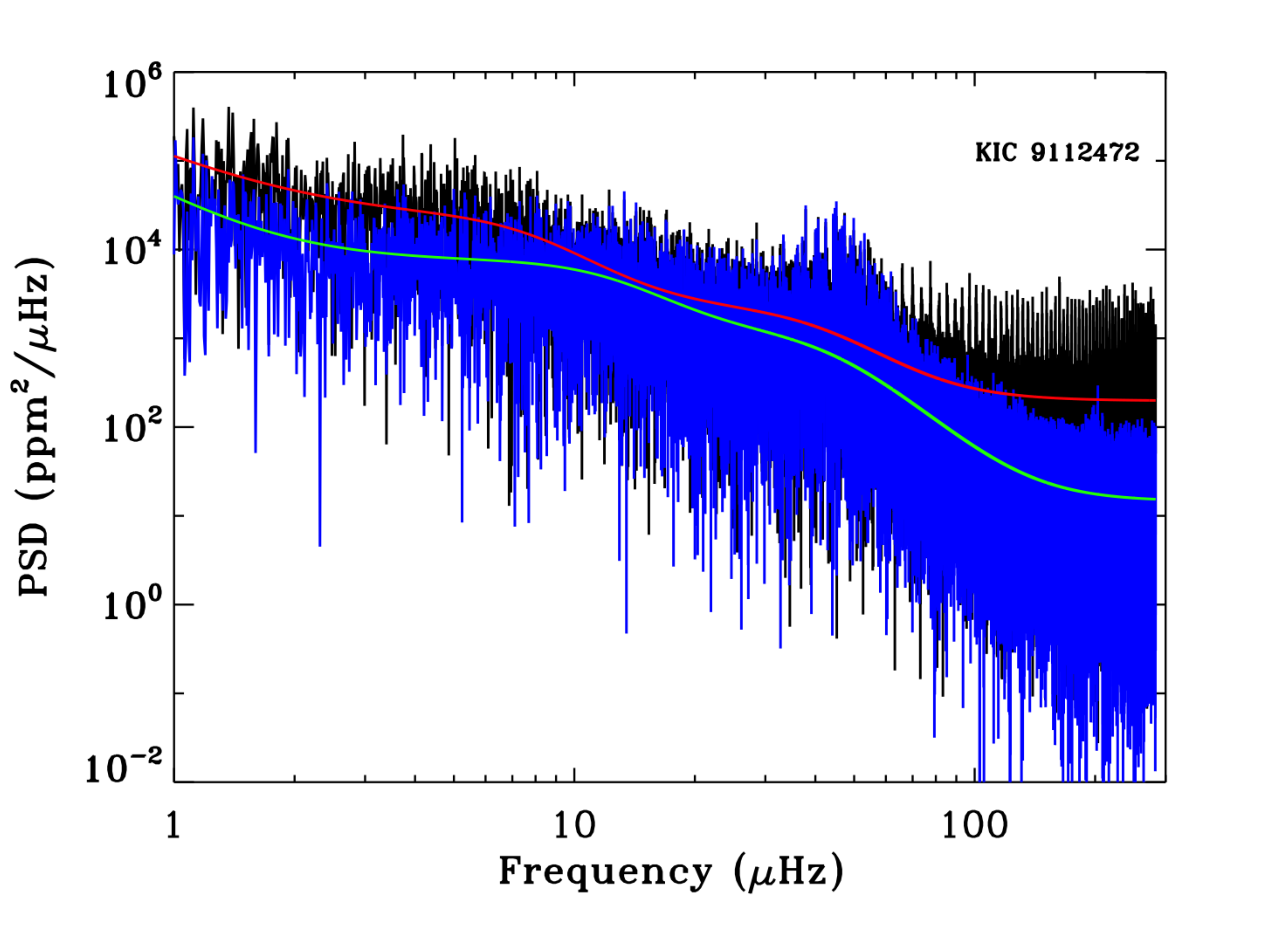}
\includegraphics[width=6cm, angle=90]{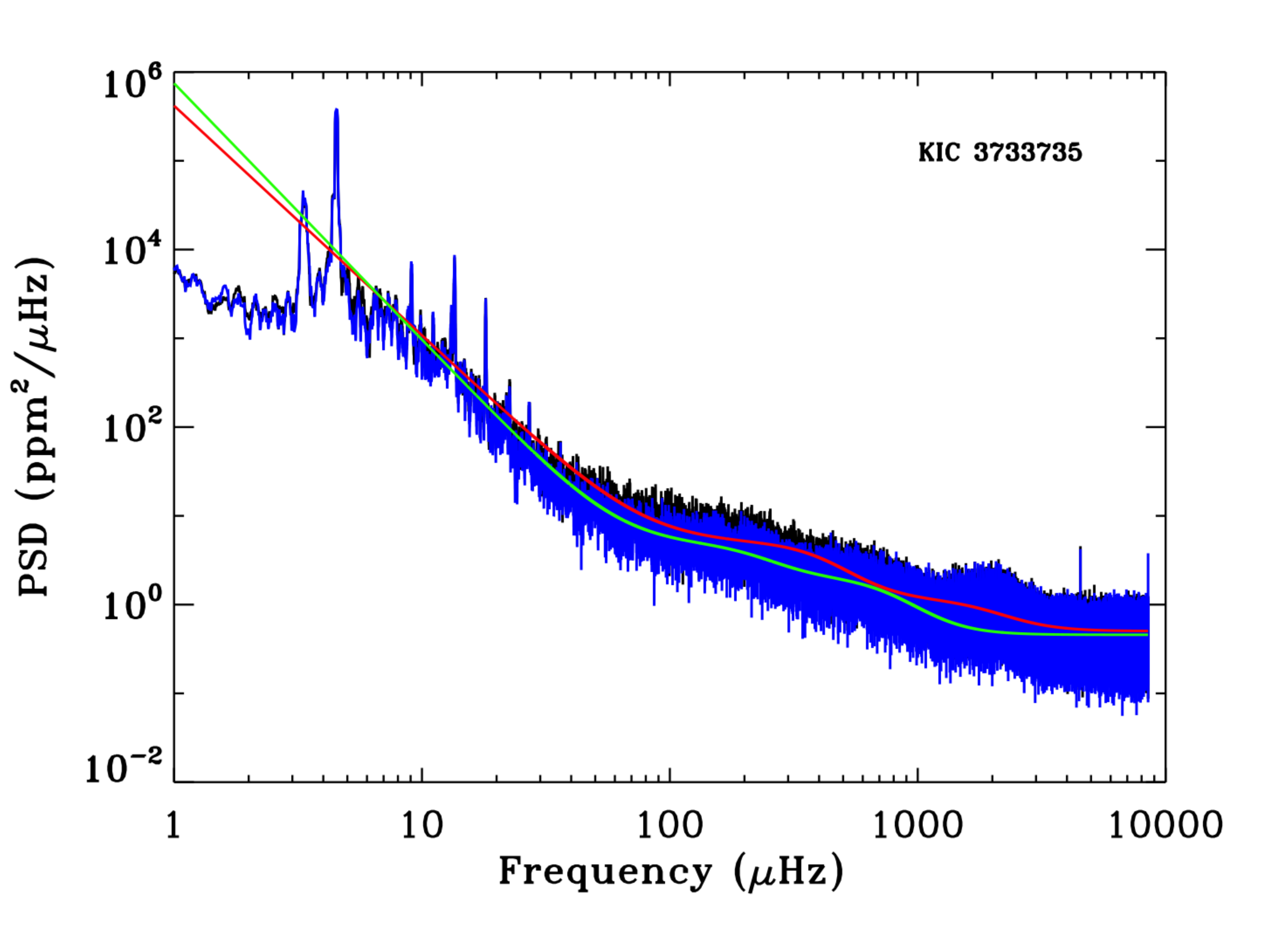}\\
\includegraphics[width=6cm, angle=90]{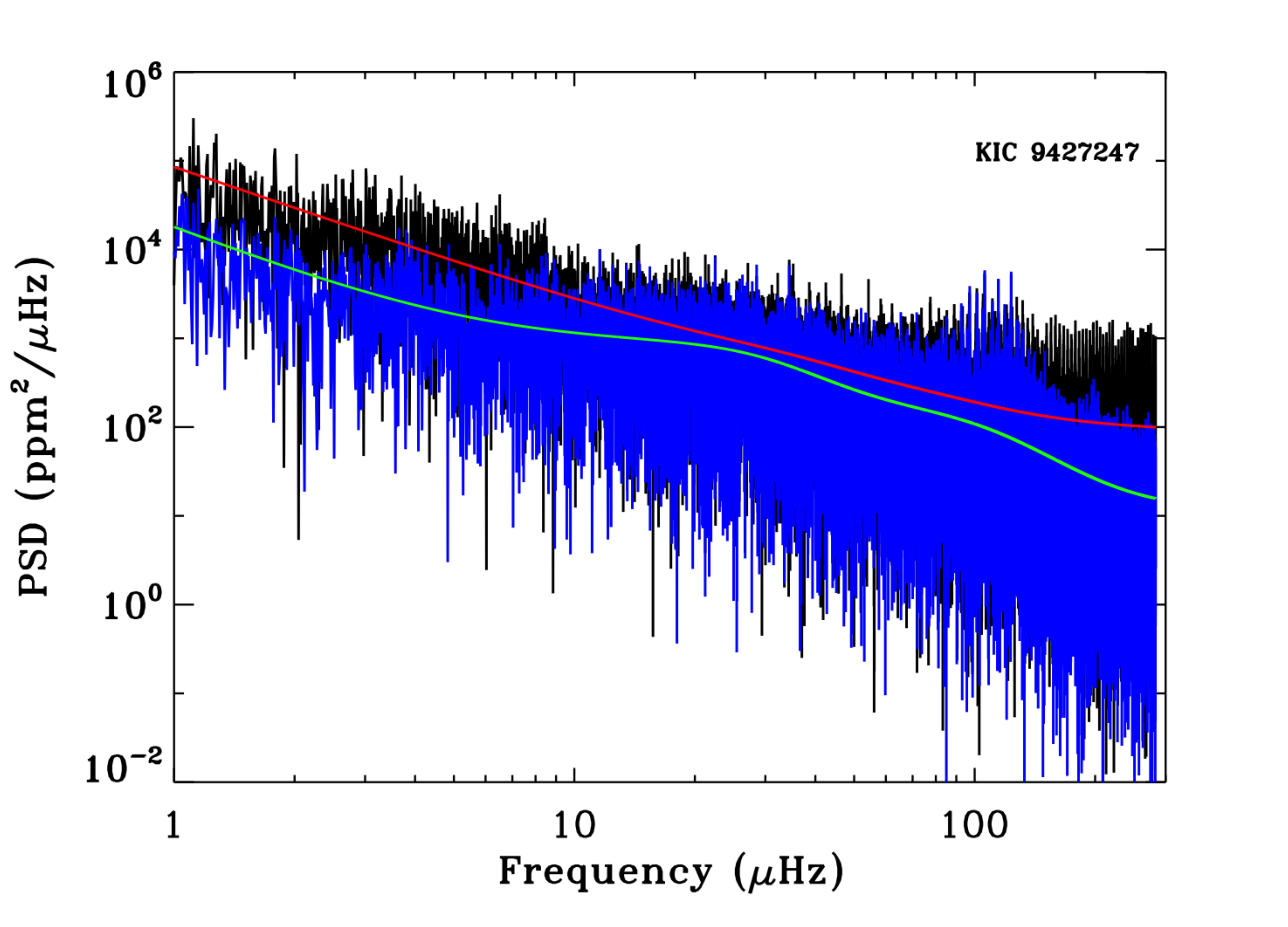}
\includegraphics[width=6cm, angle=90]{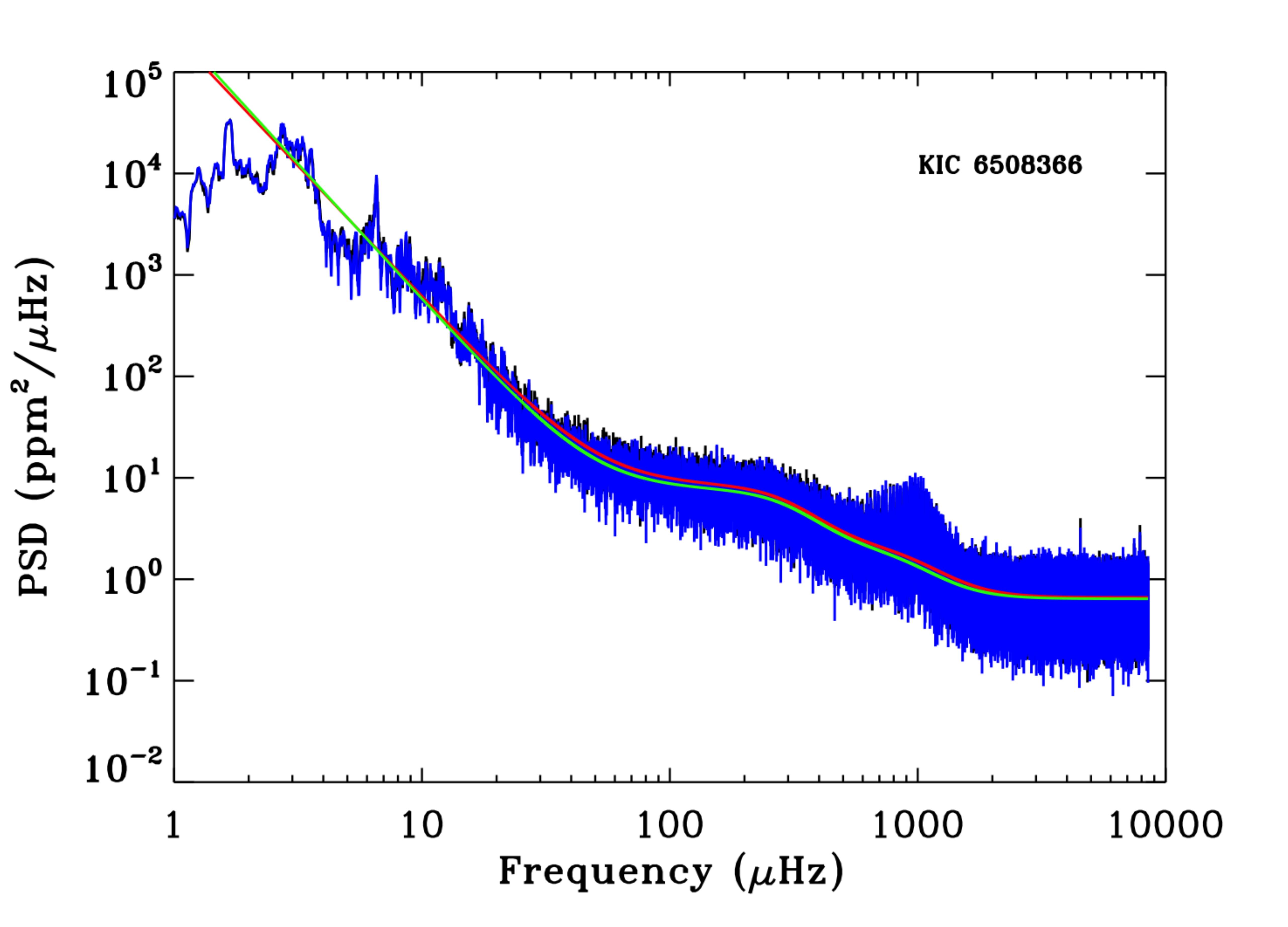}\\
\includegraphics[width=6cm, angle=90]{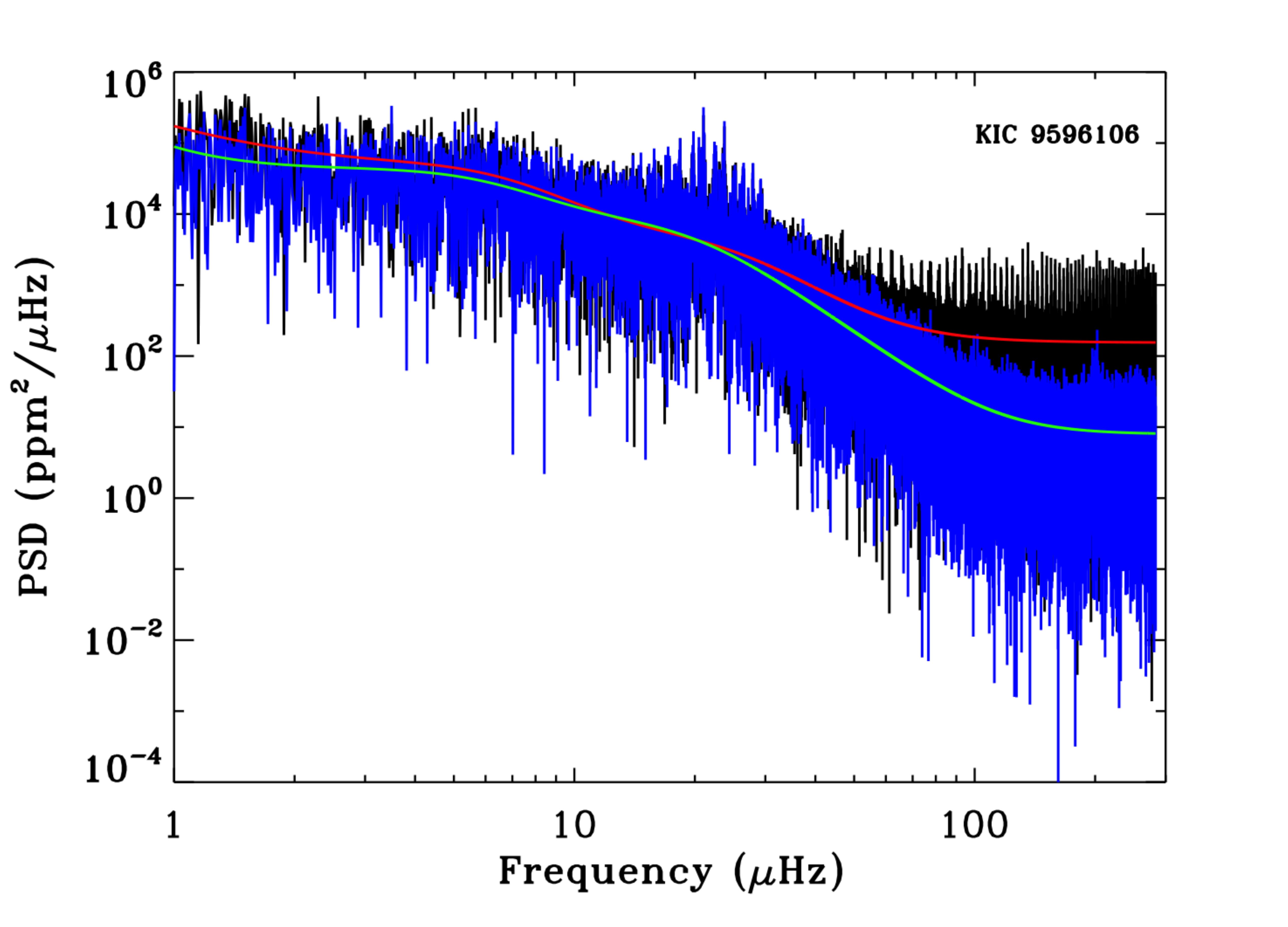}
\includegraphics[width=6cm, angle=90]{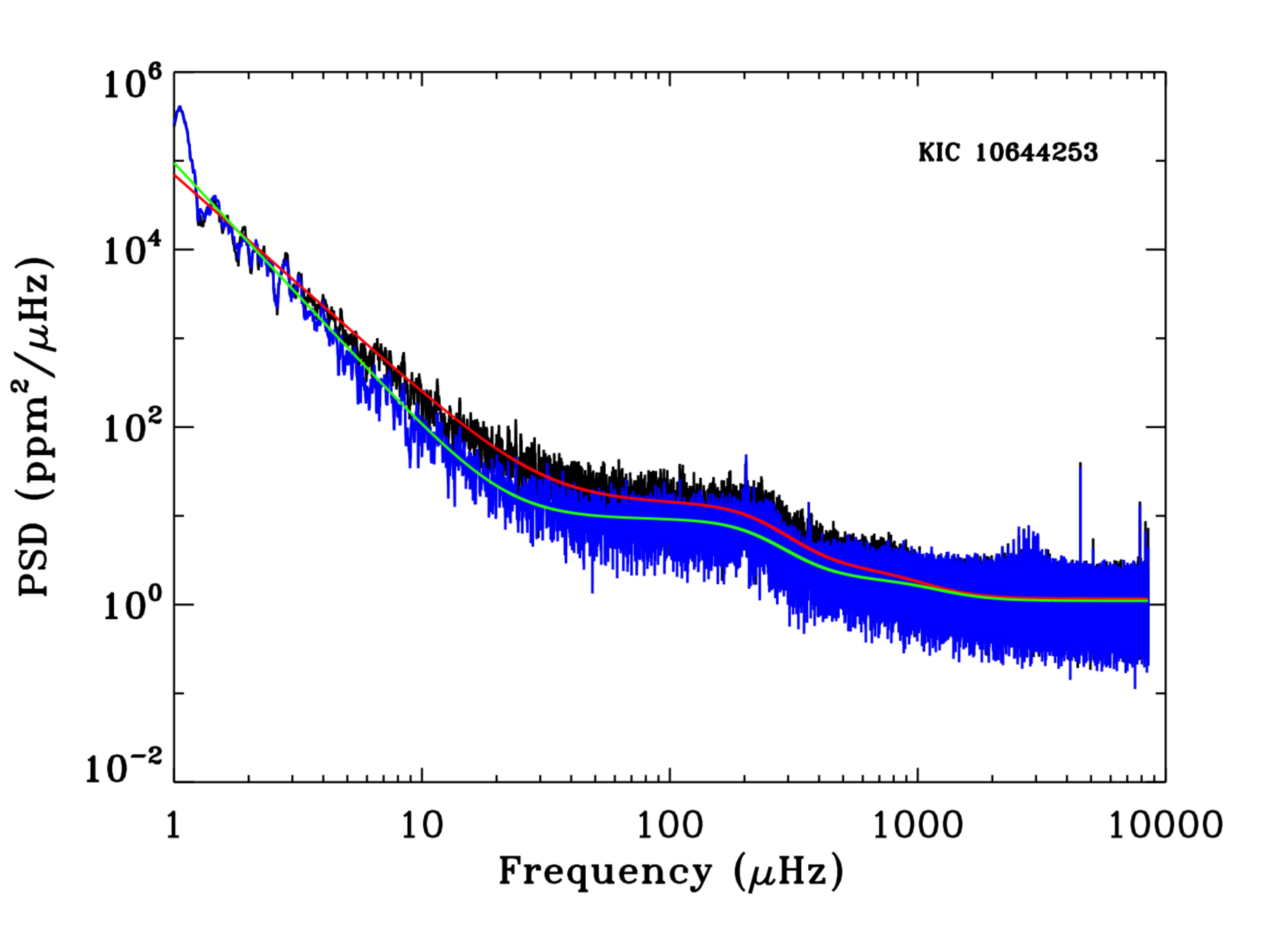}\\

\caption{Results of the background fits as describe in the text for three red giants (left panels) and three solar-like stars smoothed over 10 bins (right panels). Spectra obtained from gapped data are shown in black, while spectra after inpainting the light curves are in blue. The background fits are represented in red (resp. green) for the data with gaps (resp. inpainted data).}
\label{Back_fits}
\end{center}
\end{figure*}

{ To quantify the change on the convective parameters, we analysed a sample of red giants and main-sequence stars. We followed a similar methodology to the one described by \citet{2011ApJ...741..119M}. We fitted  a constant photon noise component dominating the spectrum at high frequency, two Harvey-like functions \citep{1985ESASP.235..199H} to measure the granulation contribution, and a power law to take into account the magnetic and rotation signatures at low frequency. For the red giants and the solar-like stars, we fixed the slopes of the Harvey-like functions to 4 (as suggested by Kallinger et al., submitted to A\&A).

To help quantify the changes, Table~\ref{back_fit} provides the background parameters and their statistical uncertainties before and after interpolating the light curves. For red giants, the white noise changes drastically with an average decrease of 90\% when the light curves are inpainted. For the granulation time scale and amplitude, there is no systematic decrease or increase but by inpainting the data they vary between 10 and 40\% for $\tau_{\rm gran}$ and 20 to 200\% for $A_{\rm gran}$.  For the solar-like stars, we observe a decrease in the white noise between 2 and 18\%, a positive or negative change in $\tau_{\rm gran}$ between 1 to 90\%, and a decrease in $A_{\rm gran}$ from 2 to 35\%. Compared to the formal uncertainties provided on these parameters, the changes are significant.

Table~\ref{back_fit} also lists the frequency of maximum power of the acoustic modes, $\nu_{\rm max}$, determined by fitting a Gaussian to the p-mode envelope. We notice that for some cases, 
there is a non-negligible change in the estimated value. This can potentially introduce a bias in the inferred stellar gravity if one uses the $\nu_{\rm{max}}$ vs. $\log g$ proportionality. For the sample of stars presented here, this can lead up to a 0.05 dex variation, i.e. 10\,$\sigma$.}




\begin{table*}[htdp]
\caption{Background parameters of six red giants and eight solar-like stars (separated by a horizontal line) for the two types of datasets (with gaps ``std'' and the inpainted data ``inp''): white noise, granulation time scale ($\tau_{\rm gran}$), granulation amplitude ($A_{\rm gran}$), and frequency of maximum power of acoustic modes ($\nu_{\rm max}$).}
\begin{center}
\begin{tabular}{rccccc}
\hline
\hline
KIC & Type & Noise ($ppm^2/\mu Hz$) & $\tau_{\rm gran}$ (ks)& $A_{\rm gran}$ ($ppm^2/\mu Hz$)& $\nu_{\rm max}$ ($\mu$Hz)\\
\hline
  8936339& std&210.90\,$\pm$\,1.01& 15.776\,$\pm$\,0.004& 
12322.55\,$\pm$\, 1415.14 & 43.8\,$\pm$\,3.3 \\
& inp& 12.81\,$\pm$\,1.01& 12.029\,$\pm$\,3.600&  7680.93\,$\pm$\,  384.00 & 37.8\,$\pm$\,1.0\\
  9112472& std&196.65\,$\pm$\,1.01& 19.762\,$\pm$\,0.002& 
20078.55\,$\pm$\, 1856.93 &47.0\,$\pm$\,2.9 \\
& inp& 14.63\,$\pm$\,1.01& 12.275\,$\pm$\,3.569&  6348.02\,$\pm$\,  368.33 & 46.5\,$\pm$\,3.2\\
  9240941& std& 58.93\,$\pm$\,1.04&  4.600\,$\pm$\,0.031& 
  317.49\,$\pm$\,   77.49 & 102.1\,$\pm$\,10.9 \\
& inp&  7.83\,$\pm$\,1.03&  5.191\,$\pm$\,1.499&   763.20\,$\pm$\,   32.29 & 109.8\,$\pm$\,4.5\\
  9364778& std&248.52\,$\pm$\,1.01& 20.445\,$\pm$\,0.002& 
22995.98\,$\pm$\, 2587.74 & 34.1\,$\pm$\,2.6\\
& inp&  4.90\,$\pm$\,1.01& 16.055\,$\pm$\,4.709& 10551.72\,$\pm$\,  629.12 & 33.0\,$\pm$\,2.2\\
  9427247& std& 84.73\,$\pm$\,1.04&  3.978\,$\pm$\,0.026& 
  232.94\,$\pm$\,   51.12 & 113.1\,$\pm$\,7.8\\
& inp& 11.25\,$\pm$\,1.04&  4.929\,$\pm$\,1.392&   711.81\,$\pm$\,   38.04 & 114.1\,$\pm$\,4.7\\
  9508595& std&103.54\,$\pm$\,1.01&  8.658\,$\pm$\,0.002& 
    26108.10\,$\pm$\, 2630.62 & 27.1\,$\pm$\,1.5\\
& inp& 20.86\,$\pm$\,1.01& 16.671\,$\pm$\,5.866& 18697.00\,$\pm$\, 1304.96 & 30.0\,$\pm$\,1.9\\
  9596106& std&154.04\,$\pm$\,1.01& 22.644\,$\pm$\,0.003& 
44854.30\,$\pm$\, 3638.83 & 20.8\,$\pm$\,1.1\\
& inp&  7.92\,$\pm$\,1.01& 24.784\,$\pm$\,7.910& 36125.23\,$\pm$\, 2570.18 & 21.1\,$\pm$\,1.2\\

\hline
  1435467& std&  0.736\,$\pm$\,0.0012&   0.478\,$\pm$\,0.004&  6.73\,$\pm$\,0.08
& 1427.6\,$\pm$\,0.1\\
& inp&  0.716\,$\pm$\,0.0011&   0.467\,$\pm$\,0.004&  5.90\,$\pm$\,0.06&
1413.8\,$\pm$\,0.1\\
  3733735& std&  0.503\,$\pm$\,0.0012&   0.367\,$\pm$\,0.005&  3.78\,$\pm$\,0.07
& 2159.6\,$\pm$\,0.2\\
& inp&  0.457\,$\pm$\,0.0007&   0.688\,$\pm$\,0.012&  2.68\,$\pm$\,0.07&
1952.2\,$\pm$\,0.2\\
  6508366& std&  0.662\,$\pm$\,0.0009&   0.493\,$\pm$\,0.008&  6.36\,$\pm$\,0.10
&  957.3\,$\pm$\,0.1\\
& inp&  0.642\,$\pm$\,0.0009&   0.489\,$\pm$\,0.008&  5.70\,$\pm$\,0.09&
 960.6\,$\pm$\,0.1\\
  7103006& std&  0.612\,$\pm$\,0.0009&   0.563\,$\pm$\,0.006&  9.96\,$\pm$\,0.17
& 1198.4\,$\pm$\,0.1\\
& inp&  0.575\,$\pm$\,0.0009&   0.551\,$\pm$\,0.006&  7.03\,$\pm$\,0.11&
1169.3\,$\pm$\,0.1\\
  7206837& std&  1.509\,$\pm$\,0.0020&   0.674\,$\pm$\,0.019&  4.59\,$\pm$\,0.11
& 1632.6\,$\pm$\,0.1\\
& inp&  1.477\,$\pm$\,0.0019&   0.641\,$\pm$\,0.019&  3.78\,$\pm$\,0.09&
1628.7\,$\pm$\,0.1\\
 10644253& std&  1.169\,$\pm$\,0.0020&   0.659\,$\pm$\,0.005& 11.45\,$\pm$\,0.16
& 2727.9\,$\pm$\,0.3\\
& inp&  1.103\,$\pm$\,0.0020&   0.668\,$\pm$\,0.005&  7.35\,$\pm$\,0.09&
2872.3\,$\pm$\,0.3\\
 12009504& std&  1.209\,$\pm$\,0.0021&   0.363\,$\pm$\,0.004&  3.37\,$\pm$\,0.04
& 1861.2\,$\pm$\,0.1\\
& inp&  0.995\,$\pm$\,0.0017&   0.354\,$\pm$\,0.004&  2.94\,$\pm$\,0.03&
1877.6\,$\pm$\,0.1\\

 \hline
\end{tabular}
\end{center}
\label{back_fit}
\end{table*}%

\begin{figure}[!htbp]
\begin{center}
\includegraphics[width=9.0cm]{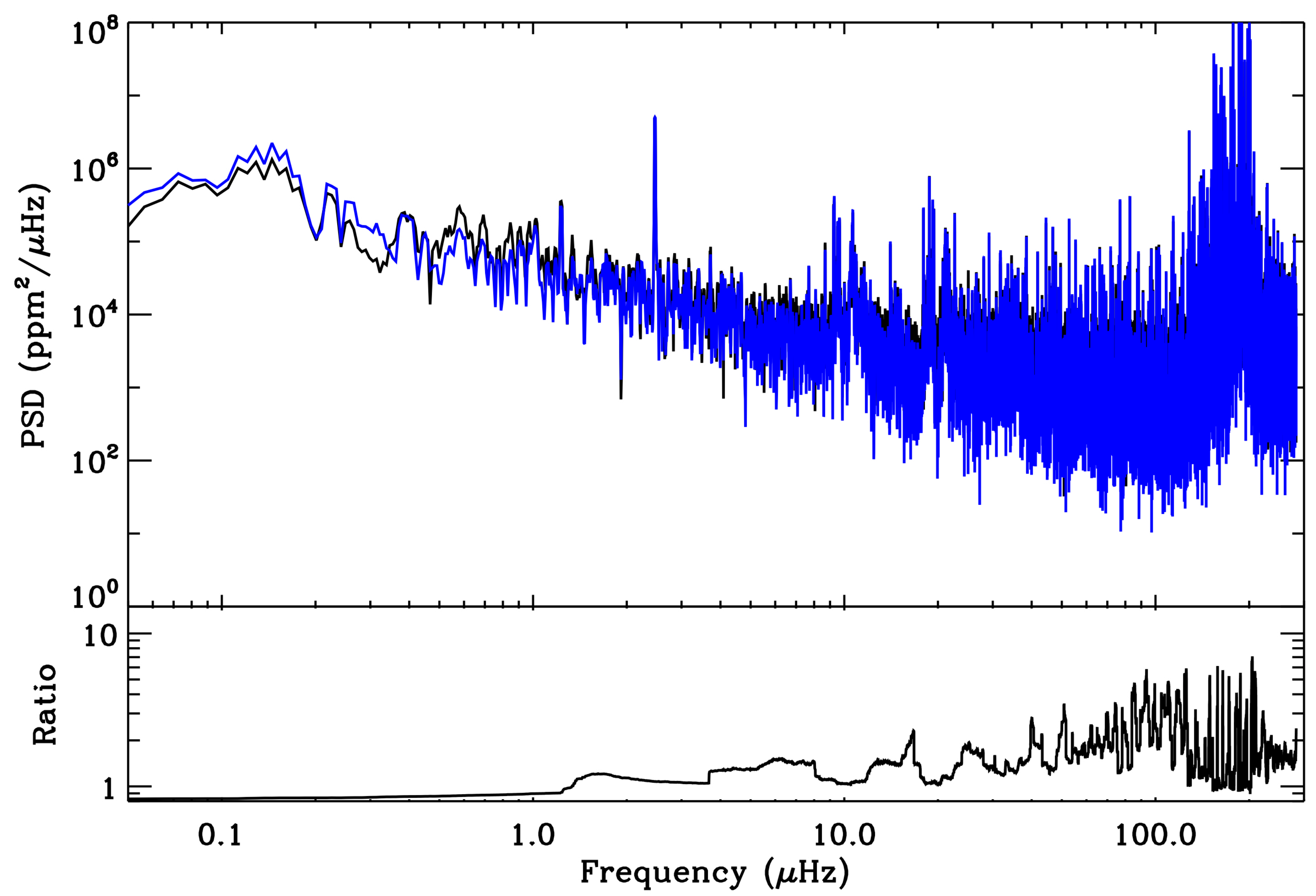}
\caption{Same than Fig.\ref{Fig_Result1} but for the classical pulsator KIC~5892969.}
\label{Fig_PSDCLAS}
\end{center}
\end{figure}

Finally, in Fig.~\ref{Fig_PSDCLAS} we show the PSD of the original (black) and inpainted (blue) light curves for the classical pulsator in the instability strip KIC~5892969. In this case, the ratio of the two spectra reveals that in the regions between the main modes, the background signal (or the ``grass level'') has been reduced by a factor between 5 to 7.

\begin{figure}[!htbp]
\begin{center}
\includegraphics[width=9.0cm]{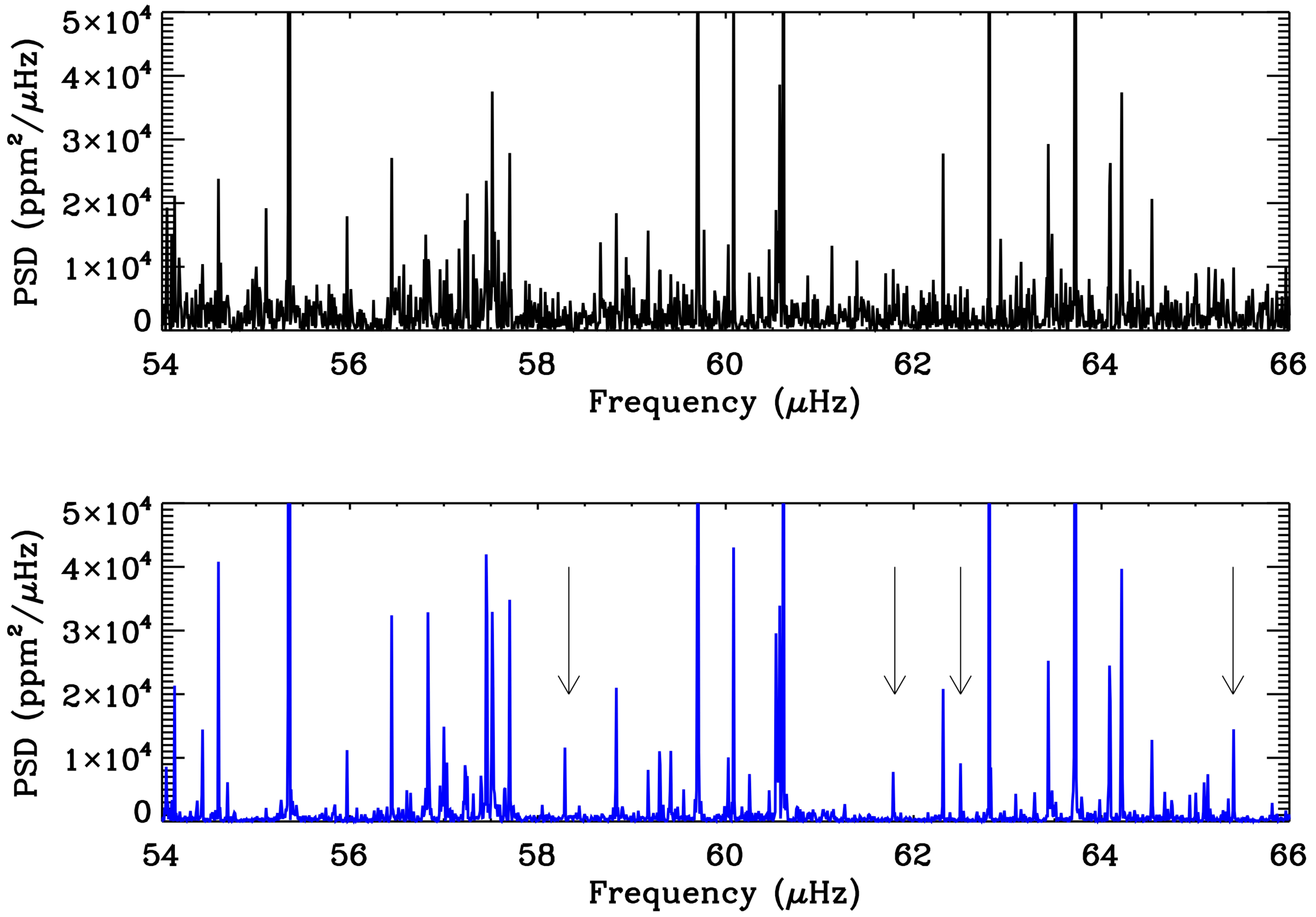}
\caption{Zoom of the PSD showed in Fig.\ref{Fig_PSDCLAS} on a region of the PSD between 54 and 66 $\mu$Hz. The top panel corresponds to the original PSD, the bottom  panel is the PSD obtained from the inpainted light  curve (in blue). Vertical arrows marked new peaks found with this analysis.}
\label{Fig_ZOOM}
\end{center}
\end{figure}

This is confirmed when we  zoom into the region between the modes. An example is shown in Fig.~\ref{Fig_ZOOM}.
The increase of the signal-to-noise ratio is clear. Most of the ``grass" of peaks have disappeared as a consequence of the reduction in the spectral window. Some new modes seem to raise (indicated by the arrows in Fig.~\ref{Fig_ZOOM}). However, a detailed analysis of the nature of these peaks is out of the scope of this paper.  

\section{Conclusions}

In this paper we have studied the nature of the regular gaps due to the nominal operations of the NASA \emph{Kepler} mission. The first sort of gaps are  due to the angular momentum desaturation of the reaction wheels producing an interruption of the data acquisition of 1 long-cadence point every three days. These gaps produce a series of harmonics at  multiples of three days (3.86 $\mu$Hz). To avoid this, and because the size of the gaps is so small, simple interpolation algorithms can be used to interpolate the missing data. The resultant interpolated PSD removes any signatures of these harmonics. 

The second type of regular gaps is due to the monthly data downlink to Earth. These gaps have a size around 0.9 day on average and thus, it is necessary to use a more complex interpolation algorithm. The effect of these gaps is more subtle and produce a trend in the background at all frequencies. 


The light curves interpolated with a $3^{rd}$ order polynomial fitting or inpainting show  a lower background level in the four categories of stars analyzed: two red giants at different stages of their evolution observed in long cadence; one young F type star observed in short cadence; and a classical pulsator in the instability strip also observed in long cadence. The resultant spectra are cleaner in all cases. In particular, for the classical pulsator for which the background level between the modes at low amplitude is highly reduced. Finally, the properties of the background also differs when the interpolated series are used. This could have a significant impact on the properties of the convective background as demonstrated for the { red giants and solar-like stars}. In this case,  the variations of the inferred granulation parameters can change { up to several tens of percent}. { We also showed that the estimation of the frequency of maximum power can be affected by the window, yielding changes in the determination of fundamental stellar parameters not in a systematic way. 
For the sample of stars presented here, this can lead up to a 0.05 dex variation, i.e. 10\,$\sigma$.
Finally, the window has an impact on the determination of the magnetic activity component of the background as we see a change of the slope at low frequency between the observations with gaps and the inpainted ones. }

For all these reasons, we recommend a proper treatment of the \emph{Kepler} regular gaps before using FFT or FFT-like transforms such as the Lomb-Scargle periodogram, in asteroseismic analyses.

\begin{acknowledgements} 
The authors wish to thank the entire \emph{Kepler} team, without whom these results would not be possible. Funding for this Discovery mission is provided by NASAÕs Science Mission Directorate. The research leading to these results has received funding from the European Research Council under the European Community's Seventh Framework Program (FP7/2007Ð2013)/ERC grant agreement ${\rm n^o}$ 227224 (PROSPERITY),  under grant agreement no. 312844 (SPACEINN), and from the Research Council of the KU Leuven under grant agreement GOA/2013/012. RAG, SM, SP, PLP, and TC have received funding from the European Community's Seventh Framework Program (FP7/2007-2013) under grant agreement no. 269194 (IRSES/ASK) and from the CNES. This work partially used data analyzed under the NASA grant NNX12AE17G. 
\end{acknowledgements} 

\appendix

\section{\emph{Kepler} timing}
\label{Ap_timing}
One of the fundamental tools in astronomy is the precise timing of astrophysical events, which has two basics sources of uncertainties: the astrophysical data that characterizes the event and the time stamp with which the event is referenced. A proper time stamp requires a reference frame (location where to measure the time) and a time standard (the way a particular clock ticks and its arbitrary zero point). Following \citet{2010PASP..122..935E}, we will designate a particular time stamp as $X_Y$, where $X$ refers to the reference frame used and $Y$ to the time standard.

The {\it Kepler} mission decided to use the $BJD_{TDB}$ as their time stamp: Barycentric Julian Date in Barycentric Dynamical Time because it is well known to be the most practical absolute time stamp for extra-terrestrial phenomena and it is ultimately limited by the properties of the target system. To be more precise, the {\it Kepler} reference frame for the time stamp was defined in relation to BJD as BKJD (Barycentric \emph{Kepler} Julian Date):  $BKJD=BJD-2454833.0$, where the offset is equal to the value of Julian Date at midday on January 1$^{st}$, 2009.

The stages and corrections performed to assign a proper time stamp to {\it Kepler} observations, can be summarized as follows:
 
\begin{itemize}
\item A local clock aboard the satellite provides the first time stamp and corresponds to the readout time for each recorded cadence (coadded and stored pixels obtained at a specific time). This time stamp is produced within 4 ms of the last pixel of the last frame and it is referred as VTC  (Vehicle Time Code).  The stability of this local oscillator clock changes with time: from quarters Q1 till Q4 it varied its rate (faster and then slower then UTC (Coordinated Universal Time)) and from Q4 till present is keeping a slower linear rate relative to UTC. At the Mission Operations Center, almost periodic resets to the clock were executed to ensure that the drift never exceeds a few seconds.  From the information extracted from KDCH, we estimate a VTC rate change of $\sim$ 2.5 s per quarter and a periodicity on the reset events of about 2.5 to 3 months. An additional correction on the timing refers to the fact that the readout time is not the same for different modules (four of them placed in the  {\it Kepler} focal plane), which has an impact when targets change location amongst them when the rotation of the spacecraft takes place  every quarter (see KDCH, for further details). 

\item The VTC time stamp of the data downloaded to Earth is corrected by the Mission Operations Center from various drifts and for leap seconds to obtain UTC. As a consequence of the previous drifts, the cadence mid-times are not evenly spaced in UTC and the cadence duration also changes (thought to be $<10^{-6}$ of the Long Cadence (LC) duration: 1766 s).

\item The final stage consists on converting UTC to TDB (Barycentric Dynamical Time) and then to correct for the motion of the spacecraft around the barycentre of the solar system.  Times corrected in this way are known as Barycentric Julian Date (BJD). The first step is accomplished by using the  well known reference JPLÕs HORIZONS ephemeris calculator. The time stamps in the data products released to the community up to Q14 (June 2012), were reported in UTC system and not in TDB, as the headers of the files state. This error was a consequence of a misinterpretation of the outputs of some routines used. Therefore an error of about 1 minute existed -due to the offset between TAI (Atomic Time) and TT (Terrestrial Time) plus the number of leap seconds- that was properly corrected in latter data releases from Q0 onwards.  The time stamp of the current released data is, therefore, $BJD_{TDB}$, or, to be precise, $BKJD_{TDB}$.

\item The main concern in relation with the {\it Kepler} analysis of its photometric time series (mainly for asteroseismology but also relevant in the context of characterization of transits) is the fact that the final cadence period (sampling time $\Delta t$) is not constant with time but varying continuously following a sinusoidal profile around the mean value.  This is an unavoidable effect associated with the reference frame chosen, the barycentre of our solar system, and due to the orbital motion of the spacecraft relative to this location. Indeed, the barycentric correction for the {\it Kepler} satellite is proportional to the semi-major axis of its orbit and the projected ecliptic latitude of the target. For the centre of the Field Of View (FOV) of {\it Kepler} this correction is of the order of $\sim\, \pm$ 211~s (see KDCH), varying periodically and changing amplitude and phase with time. Moreover, because its wide FOV, $\sim$  $10^{\rm{o}}$, the variation of the ecliptic latitude produces additional amplitude changes in the range of some $\sim \pm 200$ s for different targets in its FOV.
\end{itemize}

{ Finally, it is important to remember --as explained in KDCH-- that the accuracy of the \emph{Kepler} time stamps is not 0.5~s (related to the read-out timing), but it lies within 7~s, at 97\% confidence level, as deduced from the analysis of the pulsating star, KIC 10139564, observed simultaneously from \emph{Kepler} and from ground \citep{2012MNRAS.424.2686B}.}

\section{Effect of the \emph{Kepler} timing in the interpolation}
\label{Ap_interpol}

\begin{figure}[!htbp]   
\begin{center}
\includegraphics[width=9.0cm]{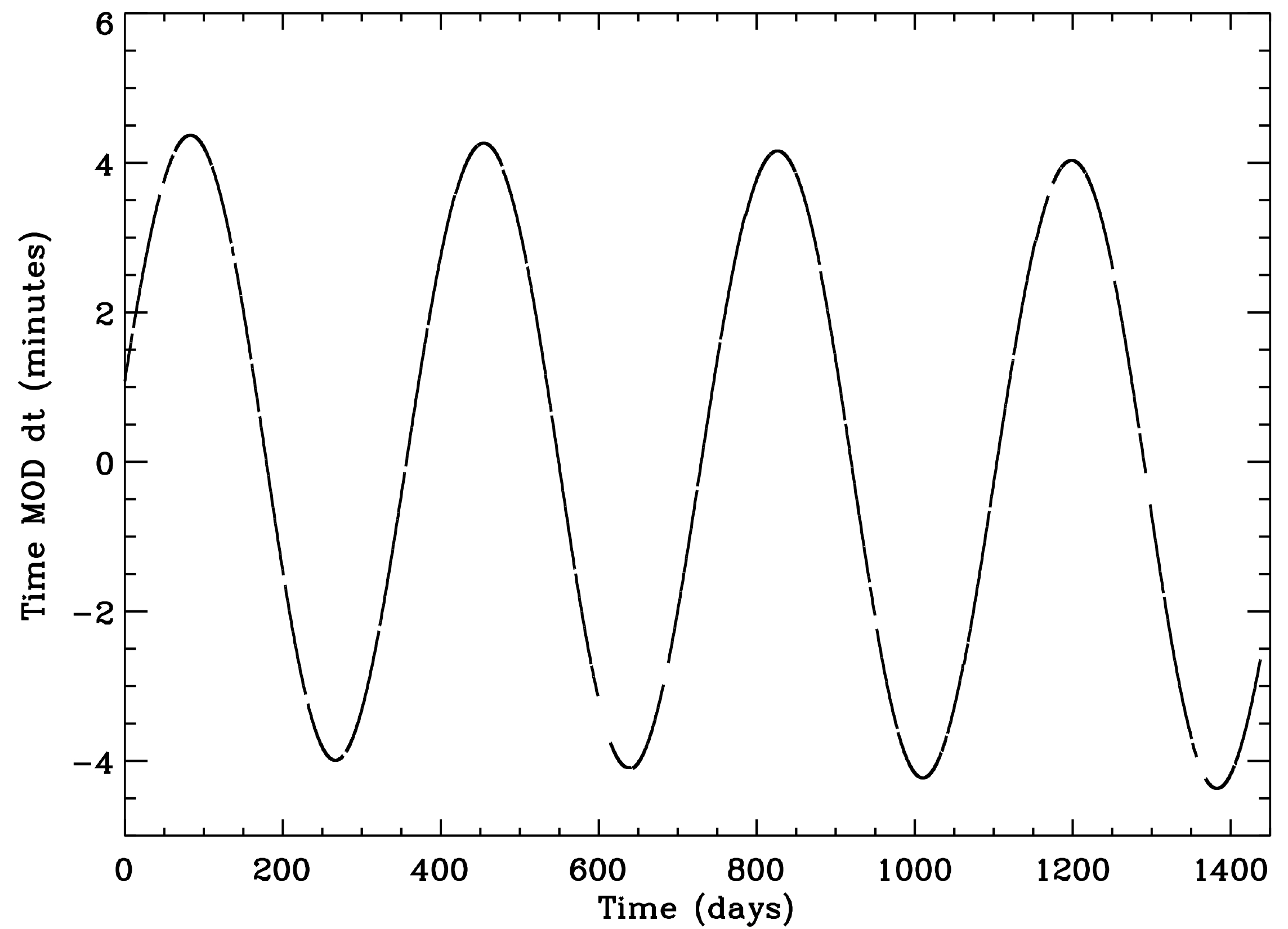}
\caption{\emph{Kepler} timing of KIC~3733735 modulo the median of its sampling rate $\Delta t$=29.424389 min.}
\label{Fig_Time}
\end{center}
\end{figure}

Because of the time stamp (reference frame and time standard) chosen for the timing of the photometric time series of {\it Kepler}, $BKJD_{TDB}$, the sampling time, $\Delta t$, is not constant but varies  (basically sinusoidaly) with time, with non-constant amplitude and phase. It also differs for different targets located in different ecliptic latitudes inside its FOV. These variations in the sampling time are  significant (as much as 10\% of the LC duration). An example can be seen in Fig.~\ref{Fig_Time}.  

\begin{figure}[!htbp]
\begin{center}
\includegraphics[width=9.0cm]{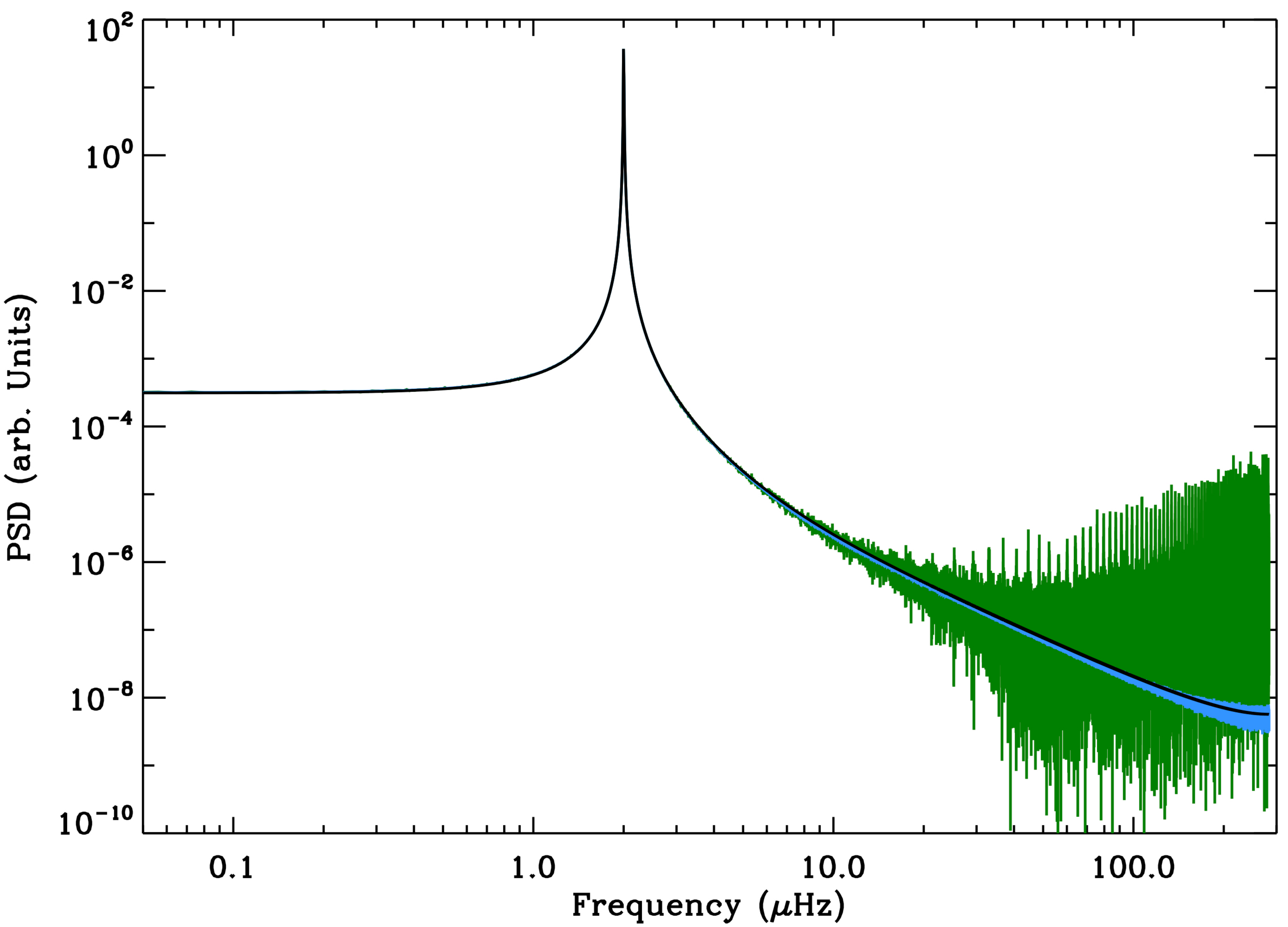}
\caption{PSD of the same simulation than in Fig.\ref{Fig_PSF} (black curve). The green curve is computed using an absolute regular grid of points in all the gaps. The light blue curve corresponds to the PSD calculated using a regular grid of points in the gaps in which the reference time for each gap is the last correct point.}
\label{Fig_PSD_mixtime}
\end{center}
\end{figure}
To properly interpolate the missing data one should take into account the sinusoidal variation of the time stamps. The first possibility consist of computing the $BKJD_{TDB}$ for each star of the field and for each quarter and use the corresponding time stamps for the interpolation. This solution is complicated and time consuming. There are three alternatives that could be easier, but not all provide good results. 
The first consists of computing an absolute regular grid of points with a sampling rate equal to the median of the original light curve, to replace the missing points. This is very fast but has the effect of mixing two different temporal sequences (a regular and an irregular). The resultant PSD is convolved by a new spectral window (see the green curve in Fig.\ref{Fig_PSD_mixtime}). The second possibility is to take a regular grid of points in the gaps but using as the reference time for each gap the one of the previous correct point. In a such way, we break the regularity in the time basis of the gaps and the increase of noise in the PSD is negligible (see blue curve in Fig.\ref{Fig_PSD_mixtime}). Finally, the third approach is the one followed in this paper and consists of converting the \emph{Kepler} timing onto a regular grid of points and then interpolating the missing points into this regular grid. An example can be seen in the blue curve of Fig.~\ref{Fig_PSF}. We have retained this option because this allows us to use the inpainting algorithm that works much faster on regular sampled time series as it takes  advantage of the Fast Fourier Transforms.

\bibliographystyle{aa}
\bibliography{./BIBLIO.bib}

\end{document}